\documentclass[12pt,a4paper]{article}

\usepackage{amsmath,amssymb,amsfonts}
\usepackage{graphicx}
\usepackage{hyperref}
\usepackage{color}
\usepackage[utf8]{inputenc}
\usepackage{physics}
\usepackage{booktabs}
\usepackage{tabularx}

\usepackage{cite}

\makeatletter
\renewcommand\section{\@startsection{section}{1}{\z@}%
  {-3.5ex \@plus -1ex \@minus -.2ex}%
  {2.3ex \@plus.2ex}%
  {\normalfont\large\bfseries\raggedright}}
\renewcommand\subsection{\@startsection{subsection}{2}{\z@}%
  {-3.25ex\@plus -1ex \@minus -.2ex}%
  {1.5ex \@plus .2ex}%
  {\normalfont\normalsize\bfseries\raggedright}}
\makeatother

\hypersetup{
    colorlinks=true,
    linkcolor=blue,
    citecolor=blue,
    urlcolor=blue
}

\newcommand{\avg}[1]{\langle#1\rangle}
\newcommand{\hc}{\text{h.c.}}

\title{Photonic Chirality for Braiding and Readout of Non-Abelian Anyons}

\author{Netzer Moriya}
\date{}
                
\begin{document}
\maketitle
                
\noindent
siOnet - Applied Modeling Research, Edison, USA\\
Corresponding author: \texttt{netzer@si-o-net.com}
\begin{abstract}
We propose a cavity-based scheme that uses photonic chirality to control braiding and read out non-Abelian anyons 
in a fractional quantum Hall platform. 
Counter-propagating cavity modes interfere with a classical reference tone to create a rotating pinning landscape 
whose direction is set by photon circulation, so that opposite photonic branches drive opposite anyon loops. 
This realizes a branch-conditioned braid operation and maps the resulting braid response onto cavity intermode coherence.
We derive the rotating pinning term and the readout relation at the effective-theory level, identify an operating window 
set by subgap driving, adiabatic transport, localization, and cavity coherence, and provide phenomenological diagnostics 
of transport locking. 
In the minimal four-anyon Ising realization, the leading signal reduces to a calibrated phase; more generally, the 
same readout structure becomes state dependent when the relative braid operator is non-scalar. 
The scheme provides a cavity route to braid-sensitive readout of non-Abelian anyons without relying on fragile 
electronic interference fringes.
\end{abstract}

\section{Introduction}
\label{sec:Introduction}
Non-Abelian anyons are a central resource for topological quantum computation, because adiabatic exchange acts by 
non-commuting holonomies on a degenerate fusion space rather than by a scalar phase \cite{Kitaev2003,Nayak2008}. 
Experimentally, however, the evidence remains indirect. Interferometric studies in gate-defined, chiral, and graphene 
devices have reported Aharonov--Bohm or phase-slip signatures consistent with anyonic 
physics \cite{Willett2013,Nakamura2019,Ghosh2025,Werkmeister2025,Kim2026}; collider and transport measurements have 
probed fractional statistics and scaling properties \cite{Bartolomei2020,Veillon2024}; and Coulomb-blockade experiments 
on Majorana islands provide complementary evidence for topological zero modes \cite{Albrecht2016,DasSarma:2015:npj}. 
Yet disorder, thermal smearing, charging effects, and edge reconstruction continue to obscure unambiguous access to 
non-Abelian braid holonomies \cite{Feldman2021}.

Foundational proposals for such measurements were developed in interferometric settings 
\cite{SternHalperin2006,BondersonKitaevShtengel2006,Bonderson2008,ChungStone2006}, and numerical studies of 
Moore--Read quasiholes mapped the braiding properties of pinned quasiholes under controlled motion \cite{Prodan2009}. 
A complementary route is to couple quantum Hall matter to microwave cavities. Experiments show that cavity vacuum 
fields can reshape quantum Hall phases \cite{Appugliese2022,Enkner2025}, while theory has explored anisotropic 
droplets \cite{Oblak2024}, synthetic fluxes \cite{Cheng2023}, moir\'e-stabilized non-Abelian phases \cite{Liu2025}, 
graviton-polariton hybridization \cite{Bacciconi2025}, and photonic braiding in synthetic platforms \cite{Zhang2022}.

What remains missing is a cavity protocol that couples a controllable photonic degree of freedom directly to the fusion 
space and reads out a braid-sensitive quantity. Here we propose such a scheme. Counter-propagating cavity modes provide 
a photonic chirality that selects the orientation of anyon transport, yielding an effective branch-conditioned braid 
operation and an interferometric cavity readout of the resulting braid response. 
The analysis is formulated at the effective-theory level: the rotating pinning landscape is derived from a dispersive 
interference response, with device-specific electrostatics absorbed into an effective coupling scale. 
The remainder of the paper develops the physical setup, the braid mechanism, the readout relation, and the associated 
operating regime.

\section{Physical Setup and Conditional Braiding Concept}
\label{sec:Physical_Setup_Conditional_Braiding_Concept}
We consider a non-Abelian fractional quantum Hall fluid, for example at $\nu=5/2$, dispersively coupled
to a superconducting ring resonator supporting counter-propagating modes $a_{\pm}$ of angular momentum $\pm m$ and resonance frequency $\omega_c$.

We encode a two-level photonic control degree of freedom in two nearly orthogonal coherent-state branches, a standard 
bosonic-encoding resource in circuit QED \cite{Leghtas2013,Mirrahimi2014,Ofek2016}; the required preparation and 
stabilization are assumed rather than modeled microscopically:
\begin{equation}
\ket{+}\equiv \ket{\alpha}_+\ket{0}_-,\qquad
\ket{-}\equiv \ket{0}_+\ket{\alpha}_-,
\label{eq:chirality_qubit_def}
\end{equation}
with mean photon number $\bar n=|\alpha|^2$. Their overlap is $\langle +|-\rangle=e^{-\bar n}$, so orthogonality 
improves with $\bar n$. The signal prefactor $\bar n e^{-\bar n}$ is maximized at $\bar n=1$, whereas the overlap 
expansion is better controlled for larger $\bar n$; one therefore works in a compromise regime where the signal remains 
appreciable and overlap corrections remain perturbative. Coherent states are used rather than Fock states because a 
nonzero field expectation value $\langle a\rangle$ is required to generate the chirality-locked potential 
$E_{\mathrm{pin}}$. The formulas below are therefore written within a branch-conditioned effective description and 
retained to leading nonvanishing order in $e^{-\bar n}$. The intended regime has $e^{-\bar n}\ll 1$, externally 
stabilized cavity amplitudes, and subleading cavity-state back-action on the anyon trajectory. Along an adiabatically 
locked trajectory in the resonant branch, the phase factor $e^{ism\phi(t)}e^{-is\delta t}$ is stationary, so 
cavity-state back-action enters only beyond the present effective description. The superposition required by 
Eq.~\eqref{eq:Psi0} is treated as part of the external bosonic-control layer; its implementation, together with the 
requirement that the reference tone $V_{\mathrm{ref}}$ not collapse the branch superposition, is deferred to 
device-specific designs.

We target Ising anyons in the Moore-Read Pfaffian state at $\nu=5/2$. The information-bearing fusion space is the 
vacuum-sector subspace of four $\sigma$ quasiholes, encoding one topological qubit. A mobile $\sigma$ anyon is 
trapped in a cavity-generated \emph{rotating} potential minimum, while three fixed $\sigma$ quasiholes are held by 
deep static pins defining the reference fusion context. Let $\Gamma$ denote the topological operation associated with 
one full counter-clockwise loop of the mobile anyon around this pinned configuration. In the explicit four-quasihole 
example below, with ordering $(1,2,3,4)$, quasihole $2$ is mobile and $\Gamma$ is one full counter-clockwise loop of 
quasihole $2$ around quasihole $3$, with quasiholes $1$ and $4$ outside the loop. The opposite cavity chirality 
realizes the inverse path $\Gamma^{-1}$.

\subsection{Effective Branch-Conditioned Braid Operation}
Let $U_{\mathrm{top}}(\Gamma)$ denote the fusion-space unitary associated with the full-loop operation $\Gamma$ in the
chosen anyon theory. Here $\Gamma$ is a full loop rather than a single elementary braid-group generator. For the
explicit Ising example below in the ordering $(1,2,3,4)$, this loop is represented by the pure braid
$\Gamma=B_2^2$, i.e.\ a double counter-clockwise exchange of quasiholes $2$ and $3$. More generally,
$U_{\mathrm{top}}(\Gamma)$ denotes the braid-word matrix associated with any other chosen full-loop path in the same
basis.

In the branch-conditioned semiclassical limit, the conditional transport is described by the effective map
\begin{equation}
\begin{aligned}
\ket{+}\otimes \ket{\psi} &\longrightarrow \ket{+}\otimes U_{\mathrm{top}}(\Gamma)\ket{\psi} \\
\ket{-}\otimes \ket{\psi} &\longrightarrow \ket{-}\otimes U_{\mathrm{top}}(\Gamma^{-1})\ket{\psi}.
\end{aligned}
\label{eq:UCB}
\end{equation}
up to corrections controlled by the coherent-branch overlap $e^{-\bar n}$. More precisely, Eq.~\eqref{eq:UCB} is a
leading-order effective map valid when the coherent branches are approximately orthogonal, the cavity amplitudes
remain stabilized during transport, and the moving anyon adiabatically follows a single rotating minimum without
significant branch-mixing back-action. For Abelian anyons, $U_{\mathrm{top}}(\Gamma)=e^{i\theta_\Gamma}$ and
Eq.~\eqref{eq:UCB} reduces to a branch-conditioned phase~\cite{Arovas1984}. For non-Abelian anyons,
$U_{\mathrm{top}}(\Gamma)$ is noncommuting, so Eq.~\eqref{eq:UCB} generically entangles the photonic branch label
with the fusion space. Because the two branches apply $U_{\mathrm{top}}(\Gamma)$ and its inverse, the interferometric
cross-term probes the corresponding relative braid operator, which in the readout convention adopted below is
$U_{\mathrm{top}}(\Gamma)^{-2}$.

To make the braid algebra concrete, choose quasiholes ordered as $(1,2,3,4)$ and the vacuum-sector basis
\begin{equation}
\ket{0_{\mathrm{L}}}
\equiv
\ket{((\sigma_1\sigma_2)_1(\sigma_3\sigma_4)_1)_1},
\qquad
\ket{1_{\mathrm{L}}}
\equiv
\ket{((\sigma_1\sigma_2)_\psi(\sigma_3\sigma_4)_\psi)_1}.
\label{eq:ising_basis_example}
\end{equation}
In one common gauge, the elementary counter-clockwise braid generators are
\begin{equation}
B_1
=
e^{-i\pi/8}
\begin{pmatrix}
1 & 0 \\
0 & i
\end{pmatrix},
\qquad
B_2
=
\frac{e^{i\pi/8}}{\sqrt{2}}
\begin{pmatrix}
1 & -i \\
-i & 1
\end{pmatrix},
\label{eq:ising_braid_generators}
\end{equation}
with $B_3=B_1$ in the same basis \cite{Nayak1996,Wu2014}.

As a concrete realization of the same physical full-loop operation, take $\Gamma$ to be the counter-clockwise monodromy of
quasihole $2$ around quasihole $3$, represented in this basis by
\begin{equation}
U_{\mathrm{top}}(\Gamma)
=
B_2^2
=
e^{-i\pi/4}
\begin{pmatrix}
0 & 1 \\
1 & 0
\end{pmatrix}.
\end{equation}
The relative operator entering the readout is therefore
\begin{equation}
U_{\mathrm{top}}(\Gamma)^{-2}
=
B_2^{-4}
=
e^{i\pi/2}\,\mathbb{1}.
\end{equation}
Accordingly, for any normalized fusion state $\ket{\psi}$ in this four-anyon vacuum-sector qubit,
\begin{equation}
\bra{\psi}U_{\mathrm{top}}(\Gamma)^{-2}\ket{\psi}
=
i.
\end{equation}
Thus the protocol realizes a genuine full loop rather than an elementary exchange, but in the minimal four-anyon
Ising setting the leading-order relative operator is scalar. The absence of state-dependent contrast is therefore a
property of this particular theory/path choice, not of the readout structure itself; Appendix
Sec.~\ref{sec:app_non_scalar_example} gives a compact example with an explicit non-scalar relative operator in the
same leading-order framework.

\section{Chirality-Locked Rotating Pinning Landscape}
\label{sec:Chirality_Locked_Rotating_Pinning}
We derive an effective rotating pinning potential at the level appropriate for a platform-agnostic theory description.
The cavity supports a chiral electric potential operator of the form
\begin{equation}
\hat V_{\mathrm{cav}}(\phi,t)
=
V_{\mathrm{zpf}}
\big(
a_+ e^{i m\phi} + a_- e^{-i m\phi}
\big)e^{-i\omega_c t} + \hc,
\label{eq:Vcav}
\end{equation}
where $V_{\mathrm{zpf}}$ is the zero-point voltage scale at the sample.
We introduce a classical reference tone $V_{\mathrm{ref}}(t)=V_d e^{-i\omega_d t}+\hc$ with $l=0$ angular structure.
In the dispersive sub-gap regime ($\hbar\omega_{c,d}\ll \Delta$), the leading time-dependent pinning term arises from 
interference between $V_{\mathrm{ref}}$ and $\hat V_{\mathrm{cav}}$ in the effective Stark shift; a compact derivation 
is given in Appendix Sec.~\ref{sec:app_derivation}.

Conditioned on cavity branch $\ket{s}$ ($s=\pm 1$), the intracavity field has $\avg{a_s}=\alpha$ and $\avg{a_{-s}}=0$.
Taking the branch expectation value of the interference term yields a classical rotating pinning landscape for the anyon coordinate $\phi$:
\begin{equation}
U_s(\phi,t) = -E_{\mathrm{pin}}\cos\!\big(m\phi - s\,\delta t - \varphi_0\big),
\quad \delta\equiv \omega_c-\omega_d,
\label{eq:Upot}
\end{equation}
where $\varphi_0$ is a controllable phase set by the reference tone and coherent-state phase.
The amplitude $E_{\mathrm{pin}}$ depends on the reference-drive amplitude, intracavity amplitude, and platform-dependent 
dispersive susceptibility; its explicit form in Appendix Eq.~\eqref{eq:SM_Epin_general} is not needed below.

Equation~\eqref{eq:Upot} implies a chirality-locked angular velocity
\begin{equation}
\Omega_s = s\,\frac{\delta}{m}.
\label{eq:Omega}
\end{equation}
The minima satisfy $m\phi-s\delta t-\varphi_0=2\pi k$, so a chosen minimum advances by $\Delta\phi = s\,\delta T/m$ 
over duration $T$, dragging a trapped anyon adiabatically.
A full $2\pi$ revolution (one braid around the pinned reference configuration) requires $\delta T = 2\pi m$, or $T=2\pi m/|\delta|$.
Loading a mobile anyon into a minimum of $U_s(\phi,t)$ and executing such a sweep realizes $\Gamma$ for $\ket{+}$ 
and $\Gamma^{-1}$ for $\ket{-}$.

At the proxy-model level, adiabatic following under noise and disorder is illustrated by dimensionless transport 
simulations with an overdamped model [Appendix Sec.~\ref{sec:app_transport}], which show locking and phase-slip suppression 
for a dragged coordinate in a moving periodic potential. This description is appropriate when the anyon coordinate is 
strongly damped and the control is slow compared with local relaxation within a minimum, but it is not a microscopic 
model of guiding-center Hall dynamics. For $m>1$, the landscape contains $m$ equivalent minima, and the requirement is 
that the anyon remain locked to one moving minimum without inter-well phase slips.

\section{Interferometric Readout of Fusion-Space Holonomy}
\label{sec:Interferometric_Readout_Fusion_Holonomy}
We initialize
\begin{equation}
\ket{\Psi_0} = \frac{1}{\sqrt{2(1+e^{-\bar{n}})}} (\ket{+}+\ket{-})\otimes \ket{\psi}_{\mathrm{top}},
\label{eq:Psi0}
\end{equation}
and, within the effective branch-conditioned description of Eq.~\eqref{eq:UCB}, obtain
\begin{equation}
\ket{\Psi_T} \approx \mathcal{N}_0\Big(\ket{+}\otimes U_{\mathrm{top}}(\Gamma)\ket{\psi}
+
\ket{-}\otimes U_{\mathrm{top}}(\Gamma)^{-1}\ket{\psi}\Big),
\label{eq:PsiT}
\end{equation}
with inherited leading-order factor $\mathcal{N}_0 = [2(1+e^{-\bar{n}})]^{-1/2}$.
The exact post-braid normalization differs by a state-dependent correction proportional to
$e^{-\bar n}\Re\!\bra{\psi}U_{\mathrm{top}}(\Gamma)^{-2}\ket{\psi}$.
Because $\bra{+}a_+^\dagger a_-\ket{-}\propto e^{-\bar n}$, this correction modifies
$\avg{a_+^\dagger a_-}$ only at order $e^{-2\bar n}$ and therefore does not affect the
leading-order readout formula Eq.~\eqref{eq:result}.
A standard heterodyne measurement can access the inter-mode coherence operator $O\equiv a_+^\dagger a_-$.
Because the two branches occupy different counter-propagating modes [Eq.~\eqref{eq:chirality_qubit_def}], 
the diagonal matrix elements vanish, $\bra{+}O\ket{+}=\bra{-}O\ket{-}=0$, and the leading nonzero contribution is the cross term
\begin{align}
\avg{a_+^\dagger a_-}
&=\bra{\Psi_T}a_+^\dagger a_-\ket{\Psi_T}\nonumber\\
&=
\frac{1}{2}\,\bra{+}a_+^\dagger a_-\ket{-}\;
\bra{\psi}U_{\mathrm{top}}(\Gamma)^{-2}\ket{\psi}
+
\mathcal O(e^{-2\bar n}),
\label{eq:result_general}
\end{align}
where we used $U_{\mathrm{top}}^\dagger(\Gamma)=U_{\mathrm{top}}(\Gamma)^{-1}$ and retained the leading nonvanishing 
order in the coherent-branch overlap. Accordingly, Eq.~\eqref{eq:result_general} is a leading-order effective readout 
relation within the same branch-conditioned semiclassical regime, not a microscopic device-resolved formula.
The photonic matrix element is computed exactly from Eq.~\eqref{eq:chirality_qubit_def}:
\begin{align}
\bra{+}a_+^\dagger a_-\ket{-}
&=
\big(\bra{\alpha}_+\bra{0}_-\big)\,a_+^\dagger a_-\,\big(\ket{0}_+\ket{\alpha}_-\big)\nonumber\\
&=
\bra{\alpha}a^\dagger\ket{0}\;\bra{0}a\ket{\alpha}\nonumber\\
&=
\Big(\alpha^*e^{-|\alpha|^2/2}\Big)\Big(\alpha e^{-|\alpha|^2/2}\Big)\nonumber\\
&= \bar n e^{-\bar n}.
\label{eq:photonic_prefactor}
\end{align}
Combining Eqs.~\eqref{eq:result_general} and \eqref{eq:photonic_prefactor} yields
\begin{equation}
\boldsymbol{\avg{a_+^\dagger a_-}}
=
\frac{1}{2}\,\mathcal{A}(\alpha)\,
\bra{\psi}U_{\mathrm{top}}(\Gamma)^{-2}\ket{\psi}
+
\mathcal O(e^{-2\bar n}),
\label{eq:result}
\end{equation}
where
\begin{equation}
\mathcal{A}(\alpha)\equiv \bar n e^{-\bar n}.
\end{equation}
Equation~\eqref{eq:result} is the leading-order readout relation within the perturbative-overlap regime described above.
Because the two chirality branches apply $U_{\mathrm{top}}(\Gamma)$ and $U_{\mathrm{top}}(\Gamma)^{-1}$, the cross-term 
probes the corresponding relative braid operator, which in the present readout convention appears through the matrix element
$\bra{\psi}U_{\mathrm{top}}(\Gamma)^{-2}\ket{\psi}$.
For suitable fusion-space states $\ket{\psi}$ and suitable full-loop paths $\Gamma$, the magnitude and phase of
$\avg{a_+^\dagger a_-}$ therefore probe braid-sensitive fusion-space overlaps. In the explicit four-quasihole Ising 
example above, $U_{\mathrm{top}}(\Gamma)^{-2}\propto \mathbb{1}$, so the signal reduces to a calibrated phase rather than 
to state-dependent contrast. This limitation is specific to that theory/path choice, not to the readout structure 
itself; Appendix Sec.~\ref{sec:app_non_scalar_example} gives a compact auxiliary non-Ising example in which 
$U_{\mathrm{top}}(\Gamma)^{-2}$ is manifestly non-scalar already in a two-dimensional fusion space.
At $\nu=5/2$, this braid-sensitive signal may help constrain or compare competing non-Abelian candidate orders when 
combined with known fusion rules, although an explicit Pfaffian/anti-Pfaffian discrimination protocol is beyond the 
scope of the present work and closely competing phases may require multiple braid trajectories. The prefactor 
$\mathcal{A}(\alpha)$ is calibratable. The non-Abelian contribution can appear as phase shifts or magnitude reduction, 
depending on the fusion state and braid; visibility suppression is not universal. Physical evolution also accumulates 
non-topological phases, $U(\Gamma) = e^{i\Phi_{\mathrm{ab}}} U_{\mathrm{top}}(\Gamma)$.

Assuming protected topological degeneracy ($\delta E \cdot T / \hbar \ll 1$, where $\delta E$ is the fusion-sector 
splitting), dynamical phases factorize as a global scalar. Since the dynamical phase is deterministic and the Abelian 
geometric phase reverses under $\Gamma \to \Gamma^{-1}$, the net scalar phase $e^{-2i\Phi_{\mathrm{ab}}}$ in 
Eq.~\eqref{eq:result} is calibratable by repeating the protocol without the enclosed anyon or with a reference Abelian 
state \cite{Willett2013,Ghosh2025}. The non-Abelian content is the fusion-space overlap 
$\bra{\psi}U_{\mathrm{top}}(\Gamma)^{-2}\ket{\psi}$, whose magnitude is state and braid dependent and can equal unity 
for eigenstates of $U_{\mathrm{top}}(\Gamma)^{-2}$ even when the statistics are non-Abelian.

To make this dependence explicit, it is convenient to express the leading-order readout for a density matrix $\rho$ over fusion 
space:
\begin{equation}
\avg{a_+^\dagger a_-}
=
\frac{1}{2}\,\mathcal{A}(\alpha)\,
\mathrm{Tr}\!\left[\rho\,U_{\mathrm{top}}(\Gamma)^{-2}\right]
+
\mathcal O(e^{-2\bar n}).
\label{eq:result_rho}
\end{equation}
For Abelian FQH quasiholes, the statistical phase from adiabatic transport can be derived via Berry 
phases~\cite{Arovas1984}; explicit non-Abelian braid-group representations for quasiholes are discussed in 
Refs.~\cite{Nayak1996,Wu2014}.

\section{Operating Regime and Robustness Criteria}
\label{sec:Operating_Regime_Robustness}
The theory above establishes the mechanism and readout without committing to specific device geometry.
Experimental implementation therefore requires parametric inequalities formulated at the effective-theory level,
because the device-dependent coupling scale $\Lambda$ has not been estimated microscopically:
(i) \emph{Sub-gap dispersive regime:} $\hbar\omega_{c,d}\ll \Delta$.
(ii) \emph{Adiabatic braid:} the braid duration $T=2\pi m/|\delta|$ must be long compared with the inverse transport gap
for motion in the moving trap, yet short compared with cavity decoherence.
A useful order-of-magnitude estimate uses the smallest relevant gap $\Delta_{\mathrm{ad}}$ encountered during transport
(for example, local trap excitations, edge-coupling gaps, or path-dependent avoided crossings):
\begin{equation}
T_{\mathrm{ad}}\ \lesssim\ T\ \ll\ \kappa^{-1},\qquad
T_{\mathrm{ad}}\ =\ \mathcal{O}\!\left(\frac{\hbar}{\Delta_{\mathrm{ad}}}\right).
\label{eq:timescale_window}
\end{equation}
Equation~\eqref{eq:timescale_window} is an operating estimate rather than a controlled adiabatic-theorem bound; the
actual condition also depends on the transport path, trap speed, local curvature, and matrix elements coupling to
excited sectors. The phenomenological overdamped diagnostics in Appendix Sec.~\ref{sec:app_transport_model} apply only
in this slow-control regime, where the rotating pinning landscape maintains localization near a single moving minimum
and the relevant issue is phase locking rather than microscopic guiding-center dynamics.
Here $\Delta_{\mathrm{ad}}$ is platform- and path-dependent. In a harmonic approximation near a moving minimum one may
estimate $\Delta_{\mathrm{ad}}\sim \hbar \omega_{\mathrm{trap}}$, with $\omega_{\mathrm{trap}}$ set by the local
curvature of the pinning landscape. Any parameter such as $m_{\mathrm{eff}}$ entering
$\omega_{\mathrm{trap}}\sim \sqrt{E_{\mathrm{pin}}/(m_{\mathrm{eff}} r_0^2)}$ is used only as a phenomenological
curvature scale, not as a claim of a microscopic quasihole band mass. $E_{\mathrm{pin}}$ primarily controls
localization against hopping.
(iii) \emph{Thermal/disorder localization:} $E_{\mathrm{pin}}$ must exceed thermal and disorder scales,
typically $E_{\mathrm{pin}} \gtrsim \mathrm{few}\times k_BT$ and larger than the disorder amplitude.
Using Eq.~\eqref{eq:SM_Epin_general}, this becomes
$2\Lambda |V_d|\,|\alpha| \gtrsim \mathrm{few}\times k_BT$.
As a purely illustrative orientation, for a GaAs/AlGaAs $\nu=5/2$ platform at $T\sim 20\,\mathrm{mK}$ one has
$k_B T \approx 1.7\,\mu\mathrm{eV}$, so localization requires $E_{\mathrm{pin}}$ of several $\mu\mathrm{eV}$.
Microwave superconducting ring resonators with $\omega_c/2\pi\sim 5$--$10\,\mathrm{GHz}$ typically have
$\kappa^{-1}\sim 1$--$10\,\mu\mathrm{s}$, setting the upper bound on $T$ in Eq.~\eqref{eq:timescale_window}.
Whether the window $T_{\mathrm{ad}}\lesssim T\ll\kappa^{-1}$ can be satisfied simultaneously depends on the achievable
$E_{\mathrm{pin}}$ and requires a device-specific estimate of $\Lambda$. Accordingly, this section should be read as a
parametric operating window rather than as a device-specific feasibility study.
(iv) \emph{Static stability:} the stationary anyon requires a pinning potential
$V_{\mathrm{static}} \gg E_{\mathrm{pin}}$.
(v) \emph{Readout visibility:} in practice, photon loss, photonic dephasing, and fusion-space decoherence give a
phenomenological envelope
$\propto e^{-(\kappa+\Gamma_\phi+\gamma_{\mathrm{top}})T}$ multiplying
Eq.~\eqref{eq:result_rho}, where $\gamma_{\mathrm{top}}$ denotes an effective topological dephasing rate.

The competition between adiabatic following and cavity coherence generically produces an optimum, as illustrated by the
dimensionless simulations in Appendix Fig.~\ref{fig:validation_SM}(c).

\section{Conclusion}
\label{sec:Conclusion}
The main theoretical claims---the effective branch-conditioned map Eq.~\eqref{eq:UCB}, the rotating pinning potential 
Eq.~\eqref{eq:Upot}, and the leading-order readout Eq.~\eqref{eq:result}---are implementation-independent at the level 
of the present effective theory. They require only a dispersive susceptibility that converts interference between a 
chiral cavity field and a reference tone into a rotating pinning potential. 

We have introduced a chirality-controlled braiding scheme in which the \emph{orientation} of anyon transport is set by 
a two-level photonic chirality degree of freedom, yielding an effective branch-conditioned braid operation and a cavity 
readout of $\mathrm{Tr}[\rho\,U_{\mathrm{top}}(\Gamma)^{-2}]$ through cross-coherence at leading non-vanishing order in 
the coherent-branch overlap. In the minimal four-anyon Ising realization worked out in the main text, this signal 
reduces to a calibrated phase; Appendix Sec.~\ref{sec:app_non_scalar_example} shows that the same readout structure 
becomes state dependent when the relative braid operator is non-scalar.

The Appendix provides detailed derivations, phenomenological proxy diagnostics of transport locking, dimensionless 
operating-window guidance, and an auxiliary non-scalar example. These diagnostics are intended as effective-theory 
consistency checks and locking proxies, not as microscopic simulations of quasihole dynamics in a specific device. By 
shifting the observable away from fragile electronic interference fringes and toward calibrated cavity coherence, the 
protocol offers a conceptually direct route to braid-sensitive access to matrix-valued non-Abelian holonomies.

\section*{Funding}
This research received no external funding.

\section*{Conflicts of Interest}
The author declares no conflicts of interest.

\section*{Data Availability}
No external datasets were used in this research. Key simulation scripts and post-processing code
for numerical validation are available from the corresponding author upon reasonable request.

\clearpage

\appendix
					  
\part*{Appendix}
\setcounter{section}{0}
\setcounter{equation}{0}
\setcounter{figure}{0}
\setcounter{table}{0}
\renewcommand{\theequation}{A\arabic{equation}}
\renewcommand{\thefigure}{A\arabic{figure}}
\renewcommand{\thetable}{A\arabic{table}}
\renewcommand{\thesection}{A\arabic{section}}
\renewcommand{\thesubsection}{\thesection.\arabic{subsection}}
\renewcommand{\theHequation}{app.eq.\arabic{equation}}
\renewcommand{\theHfigure}{app.fig.\arabic{figure}}
\renewcommand{\theHtable}{app.tab.\arabic{table}}
\renewcommand{\theHsection}{app.sec.\arabic{section}}
\renewcommand{\theHsubsection}{app.subsec.\arabic{section}.\arabic{subsection}}

\addcontentsline{toc}{section}{Appendix}
\label{app:Appendix}

\section{Derivation of the Chirality-Locked Rotating Pinning Landscape}
\label{sec:app_derivation}
This section derives Eq.~\eqref{eq:Upot} in a form that makes the operator content and scaling explicit.

\subsection{Effective dispersive coupling and the interference term}

We model the cavity electric potential at the sample plane as
\begin{equation}
\hat V_{\mathrm{cav}}(\phi,t)
=
V_{\mathrm{zpf}}
\Big(
a_+ e^{i m\phi} + a_- e^{-i m\phi}
\Big)e^{-i\omega_c t}
+\hc,
\label{eq:SM_Vcav}
\end{equation}
where $V_{\mathrm{zpf}}$ is the zero-point voltage scale at the sample, and the reference tone as
\begin{equation}
V_{\mathrm{ref}}(t)=V_d e^{-i\omega_d t}+\hc,
\label{eq:SM_Vref}
\end{equation}
with $l=0$ angular structure. In the sub-gap dispersive regime $\hbar\omega_{c,d}\ll \Delta$, the leading effective
energy shift of the gapped fluid is quadratic in the total potential and may be written schematically as
\begin{equation}
H_{\mathrm{eff}}
=
-\int d^2r\ \chi(\mathbf r)\ :\!|\hat V_{\mathrm{cav}}(\mathbf r,t)+V_{\mathrm{ref}}(t)|^2\!:\ ,
\label{eq:SM_Heff}
\end{equation}
where $\chi(\mathbf r)$ is an effective static susceptibility, taken to be a real kernel that depends on the platform
and on the relevant virtual excitations of the gapped fluid. Equation~\eqref{eq:SM_Heff} states only that the dominant
response is dispersive and even in the electric field; its microscopic justification and the detailed structure of
$\chi$ are system-dependent and are not needed for the kinematic chirality-locking result. In the intended sub-gap
regime, the dominant contribution to $\chi(\mathbf r)$ is expected to come from virtual electronic excitations across
the lowest available gapped sector of the quantum Hall fluid, such as the many-body bulk gap within the relevant
Landau level. Higher-energy processes, including inter-Landau-level virtual transitions when present, may be absorbed
into the same static effective kernel within the present dispersive description.

Expanding Eq.~\eqref{eq:SM_Heff}, the time-dependent rotating contribution arises from the interference cross-term
\begin{equation}
H_{\mathrm{int}}
=
-2\int d^2r\ \chi(\mathbf r)\ \mathrm{Re}\!\left[V_{\mathrm{ref}}^*(t)\,\hat V_{\mathrm{cav}}(\mathbf r,t)\right].
\label{eq:SM_Hint_def}
\end{equation}
This term is linear in the cavity field operator and therefore produces a pinning landscape proportional to the
coherent amplitude $\avg{a_s}$ in the occupied branch; it is not number-diagonal by itself. Number-diagonal AC-Stark
contributions arise from the $|\hat V_{\mathrm{cav}}|^2$ term, but in the present single-tone setting they are
time-independent and therefore do not generate rotation at $\delta$.

\subsection{Rotating-wave approximation and emergence of a rotating cosine}

Substituting Eqs.~\eqref{eq:SM_Vcav} and \eqref{eq:SM_Vref} into Eq.~\eqref{eq:SM_Hint_def} and retaining resonant
terms near the detuning $\delta=\omega_c-\omega_d$ gives
\begin{equation}
H_{\mathrm{int}}
\simeq
-\int d^2r\ \chi(\mathbf r)\ V_{\mathrm{zpf}}
\Big[
V_d^*
\big(
a_+ e^{i m\phi} + a_- e^{-i m\phi}
\big)e^{-i\delta t}
+\hc
\Big].
\label{eq:SM_Hint_RWA}
\end{equation}
For a localized quasiparticle pinned near a radius whose slow collective coordinate is the polar angle $\phi$, we
project the spatial integral onto an effective coupling constant $\Lambda$, set by the geometry and by $\chi$, and
write
\begin{equation}
H_{\mathrm{int}}
\simeq
-\Lambda
\Big[
V_d^*
\big(
a_+ e^{i m\phi} + a_- e^{-i m\phi}
\big)e^{-i\delta t}
+\hc
\Big],
\label{eq:SM_Hint_projected}
\end{equation}
where $\Lambda$ has dimensions of energy per voltage and absorbs the geometric and susceptibility factors, consistent
with Eq.~\eqref{eq:SM_Hint_RWA}. At this point the derivation passes from an explicit spatial response kernel to a
coarse-grained effective overlap parameter, so the result should be read as an effective scaling reduction rather than
as a fully microscopic electrostatic calculation. In practice, $\Lambda$ collects the cavity-field geometry, the
overlap of the rotating field pattern with the localized quasihole, and the effective susceptibility kernel
$\chi(\mathbf r)$. This projection is sufficient to derive the chirality-locked rotating potential and the scaling
$E_{\mathrm{pin}} = 2\Lambda |V_d| |\alpha|$, but it does not by itself determine the absolute pinning strength for a
specific device. A quantitative estimate of $\Lambda$ therefore requires an implementation-specific microscopic or
electrostatic model.

Conditioning on a chirality branch $\ket{s}$ with $\avg{a_s}=\alpha$ and $\avg{a_{-s}}=0$, the effective classical
pinning potential acting on the anyon coordinate is obtained by taking the branch expectation value:
\begin{align}
U_s(\phi,t)
&\equiv
\avg{H_{\mathrm{int}}}_{\ket{s}}
\simeq
-\Lambda\Big[
V_d^*\,\alpha\,e^{i s m\phi}\,e^{-i\delta t}+\hc
\Big]\nonumber\\
&=
-2\Lambda |V_d|\,|\alpha|\,
\cos\!\big(m\phi - s\,\delta t-\varphi_0\big),
\label{eq:SM_U_rotating}
\end{align}
with $\varphi_0\equiv \arg(V_d^*\alpha)$. Identifying
\begin{equation}
E_{\mathrm{pin}}\equiv 2\Lambda |V_d|\,|\alpha|
\label{eq:SM_Epin_general}
\end{equation}
yields Eq.~\eqref{eq:Upot} in the main text. The factor of $2$ follows from combining
Eq.~\eqref{eq:SM_Hint_def}, where the interaction is written as $-2\,\mathrm{Re}[\cdots]$, with
$\mathrm{Re}[z]=(z+z^*)/2$. The derivation therefore makes explicit that the rotating pinning term scales as
$E_{\mathrm{pin}}\propto |V_d|\,|\alpha|$, that is, with the coherent amplitude rather than with $\bar n$ alone in
the single-tone interference mechanism.

\section{Photonic prefactor for cross-coherence readout}

The photonic matrix element entering Eqs.~\eqref{eq:photonic_prefactor}--\eqref{eq:result} follows directly from the 
branch definitions $\ket{+}=\ket{\alpha}_+\ket{0}_-$ and $\ket{-}=\ket{0}_+\ket{\alpha}_-$:
\begin{align}
\bra{+}a_+^\dagger a_-\ket{-}
&=
\bra{\alpha}_+\bra{0}_- a_+^\dagger a_- \ket{0}_+\ket{\alpha}_-\nonumber\\
&=
\bra{\alpha}a^\dagger\ket{0}\;\bra{0}a\ket{\alpha}\nonumber\\
&=
\left(\bra{0}a\ket{\alpha}\right)^*\left(\bra{0}a\ket{\alpha}\right)\nonumber\\
&=
|\alpha|^2 e^{-|\alpha|^2} = \bar n e^{-\bar n}.
\end{align}

\section{Explicit non-scalar example of state-dependent contrast}
\label{sec:app_non_scalar_example}

To show that the scalar outcome of the minimal four-anyon Ising example is not intrinsic to the readout principle, we 
give the following auxiliary example; the physical target of the main text remains the Ising/$\nu=5/2$ setting. Let 
three Fibonacci anyons $\tau$ with total topological charge $\tau$ span a two-dimensional fusion space in a standard 
convention~\cite{Nayak2008}. In the basis
\begin{equation}
\ket{0_L}\equiv \ket{((\tau_1\tau_2)_1\tau_3)_\tau},
\qquad
\ket{1_L}\equiv \ket{((\tau_1\tau_2)_\tau\tau_3)_\tau},
\end{equation}
the elementary counter-clockwise exchange of anyons $1$ and $2$ is
\begin{equation}
B_{1}^{(\mathrm{Fib})}
=
\begin{pmatrix}
e^{-4\pi i/5} & 0\\
0 & e^{3\pi i/5}
\end{pmatrix}.
\end{equation}
Let the full-loop path $\Gamma_{\mathrm{Fib}}$ be the counter-clockwise monodromy of anyon $1$ around anyon $2$, so that
\begin{equation}
U_{\mathrm{top}}(\Gamma_{\mathrm{Fib}})
=
\left(B_{1}^{(\mathrm{Fib})}\right)^2.
\end{equation}
The relative operator entering the leading-order readout is then
\begin{equation}
U_{\mathrm{top}}(\Gamma_{\mathrm{Fib}})^{-2}
=
\left(B_{1}^{(\mathrm{Fib})}\right)^{-4}
=
\begin{pmatrix}
e^{6\pi i/5} & 0\\
0 & e^{-2\pi i/5}
\end{pmatrix},
\end{equation}
which is manifestly non-scalar. Hence the leading-order signal
\begin{equation}
\avg{a_+^\dagger a_-}
=
\frac{1}{2}\,\mathcal{A}(\alpha)\,
\mathrm{Tr}\!\left[\rho\,U_{\mathrm{top}}(\Gamma_{\mathrm{Fib}})^{-2}\right]
+
\mathcal{O}(e^{-2\bar n})
\end{equation}
is already state dependent in a two-dimensional fusion space. This example shows that the scalar outcome of the minimal 
four-anyon Ising realization is a property of that specific theory/path choice, not of the readout structure itself.

\section{Numerical diagnostics of transport locking (dimensionless proxy)}
\label{sec:app_transport}
The simulations serve only as phenomenological proxy diagnostics of
(i) a broad locking regime for a dragged coordinate in a rotating cosine trap with static disorder and noise, and
(ii) an optimal compromise between adiabatic following and cavity coherence.
They are not microscopic simulations of quasihole guiding-center dynamics; rather, they represent an effective classical
locking model appropriate to a strongly dissipative limit in which the slow coordinate of a trapped anyon is
approximated as overdamped in the co-rotating pinning landscape.

\subsection{Minimal stochastic model}
\label{sec:app_transport_model}
We simulate overdamped stochastic dynamics for an angular coordinate $\phi(t)$, treated as an effective coarse-grained
locking coordinate, in a rotating pinning landscape plus static disorder:
\begin{equation}
U_s(\phi,t)= -E_{\mathrm{pin}}\cos\!\big(\phi-s\varphi(t)\big)\;-\;U_{\mathrm{dis}}\cos(\phi+\theta),
\label{eq:SM_U_MC}
\end{equation}
where $\varphi(t)=2\pi t/T$ and the disorder phase $\theta$ is sampled uniformly on $[0,2\pi)$ for each trajectory.
The dynamics follow the overdamped Langevin equation
\begin{equation}
\dd\phi = -\mu\,\partial_\phi U_s(\phi,t)\,\dd t + \sqrt{2D\,\dd t}\;\eta(t),
\label{eq:SM_Langevin}
\end{equation}
where $D=\mu k_BT_{\mathrm{eff}}$ and $\eta(t)$ is Gaussian white noise. For $m>1$, the dynamics reduce to the same form
after the rescaling $\phi' = m\phi$, preserving the locking physics.

This model is a dimensionless proxy for the existence of a locking regime; it does not encode full guiding-center Hall
dynamics. The one-dimensional reduction is motivated by the assumption that the rotating pinning landscape localizes the
quasihole near a preferred radius $r_0$, so radial fluctuations are gapped by the local trap curvature and the slow
collective coordinate is the azimuthal angle $\phi(t)$. In the underlying fractional quantum Hall problem, guiding-center
motion in a strong magnetic field is first-order and two-dimensional, so the overdamped Langevin equation is not a
microscopic equation of motion but a coarse-grained model for phase locking, noise tolerance, and loss of adiabatic
tracking. It is therefore appropriate only for diagnosing whether a well-localized quasihole remains locked to a single
moving minimum under slow driving and noise, not for extracting microscopic transport coefficients, Berry-curvature
corrections, or coherent oscillatory dynamics.

For fixed $\mu$ and $k_BT_{\mathrm{eff}}$, a sweep in $(v,E_{\mathrm{pin}})$ is equivalent, up to axis rescaling, to a
sweep in the dimensionless ratios $(x,y)$ used in Fig.~\ref{fig:validation_SM}(b):
(i) the depth-to-noise ratio
\begin{equation}
y \equiv \frac{E_{\mathrm{pin}}}{k_BT_{\mathrm{eff}}},
\label{eq:SM_y_def}
\end{equation}
and (ii) the reduced angular speed,
\begin{equation}
x \equiv \frac{v}{\mu E_{\mathrm{pin}}},\qquad v\equiv \frac{2\pi}{T}.
\label{eq:SM_x_def}
\end{equation}
Large $y$ suppresses thermally assisted inter-minimum hopping, whereas small $x$ favors adiabatic following of the
moving minimum.

\subsection{Phase-slip locking success criterion}
\label{sec:app_phase_slip}
For diagnostics we also record the net winding number at the protocol end,
\begin{equation}
W_s \equiv \frac{\phi(T)-\phi(0)}{2\pi},
\label{eq:SM_winding}
\end{equation}
but the success criterion used for Fig.~\ref{fig:validation_SM}(b) is the stricter phase-slip locking test in the
co-rotating frame. Define
\begin{equation}
\psi_s(t)\equiv \phi(t)-s\,\varphi(t),
\label{eq:SM_psi_def}
\end{equation}
so that the rotating trap term in Eq.~\eqref{eq:SM_U_MC} becomes $-E_{\mathrm{pin}}\cos\psi_s$.
Adjacent minima are separated by barriers at $|\psi_s|=\pi$. A phase slip is therefore any crossing of this barrier
during the protocol, and a trajectory is declared successful if
\begin{equation}
\max_{t\in[0,T]}|\psi_s(t)| < \pi .
\label{eq:SM_success}
\end{equation}
From ensembles of independent disorder and noise realizations we estimate $P_{\mathrm{succ}}^{\pm}$ for $s=\pm1$ and
the symmetrized probability $P_{\mathrm{succ}}=(P_{\mathrm{succ}}^{+}+P_{\mathrm{succ}}^{-})/2$. This criterion directly
tests whether the coordinate remains trapped in a single moving minimum and is more stringent than endpoint-only winding
tolerances.

\subsection{Visibility envelope illustration}
\label{sec:app_visibility}
A simple envelope for the experimentally expected coherence magnitude is
\begin{equation}
\begin{aligned}
\big|\langle a_+^\dagger a_-\rangle\big|
&\approx
\frac{1}{2}\,\mathcal{A}(\alpha)\,P_{\mathrm{succ}}(E_{\mathrm{pin}},T) \\
&\quad \times \exp\!\big[-(\kappa+\Gamma_\phi)T\big]\,
\big|\mathrm{Tr}[\rho\,U_{\mathrm{top}}(\Gamma)^{-2}]\big|.
\end{aligned}
\label{eq:SM_visibility}
\end{equation}
For the illustrative proxy plots we omit the additional topological-dephasing factor $\exp(-\gamma_{\mathrm{top}}T)$
that appears in the main-text phenomenological envelope; it can be reinstated as an additional exponential factor. It is
convenient to factor out the photonic prefactor $\mathcal{A}(\alpha)$ and define the normalized coherence proxy
\begin{equation}
\begin{aligned}
\mathcal{C}(T) &\equiv \frac{2\,\big|\langle a_+^\dagger a_-\rangle\big|}{\mathcal{A}(\alpha)} \\
&= P_{\mathrm{succ}}(E_{\mathrm{pin}},T)\,\exp\!\big[-(\kappa+\Gamma_\phi)T\big] \\
&\quad \times \big|\mathrm{Tr}[\rho\,U_{\mathrm{top}}(\Gamma)^{-2}]\big|.
\end{aligned}
\label{eq:SM_C_def}
\end{equation}
Panel~(c) of Fig.~\ref{fig:validation_SM} plots $\mathcal{C}(T)$; the absolute expected coherence is recovered by
multiplying by $\mathcal{A}(\alpha)/2$. The normalized proxy illustrates the generic existence of an optimal $T$
arising from the competition between increasing $P_{\mathrm{succ}}$ and decreasing cavity coherence.

\subsection{Appendix diagnostic figure and metrics}

\begin{figure*}[!t]
\centering
\includegraphics[width=0.32\textwidth]{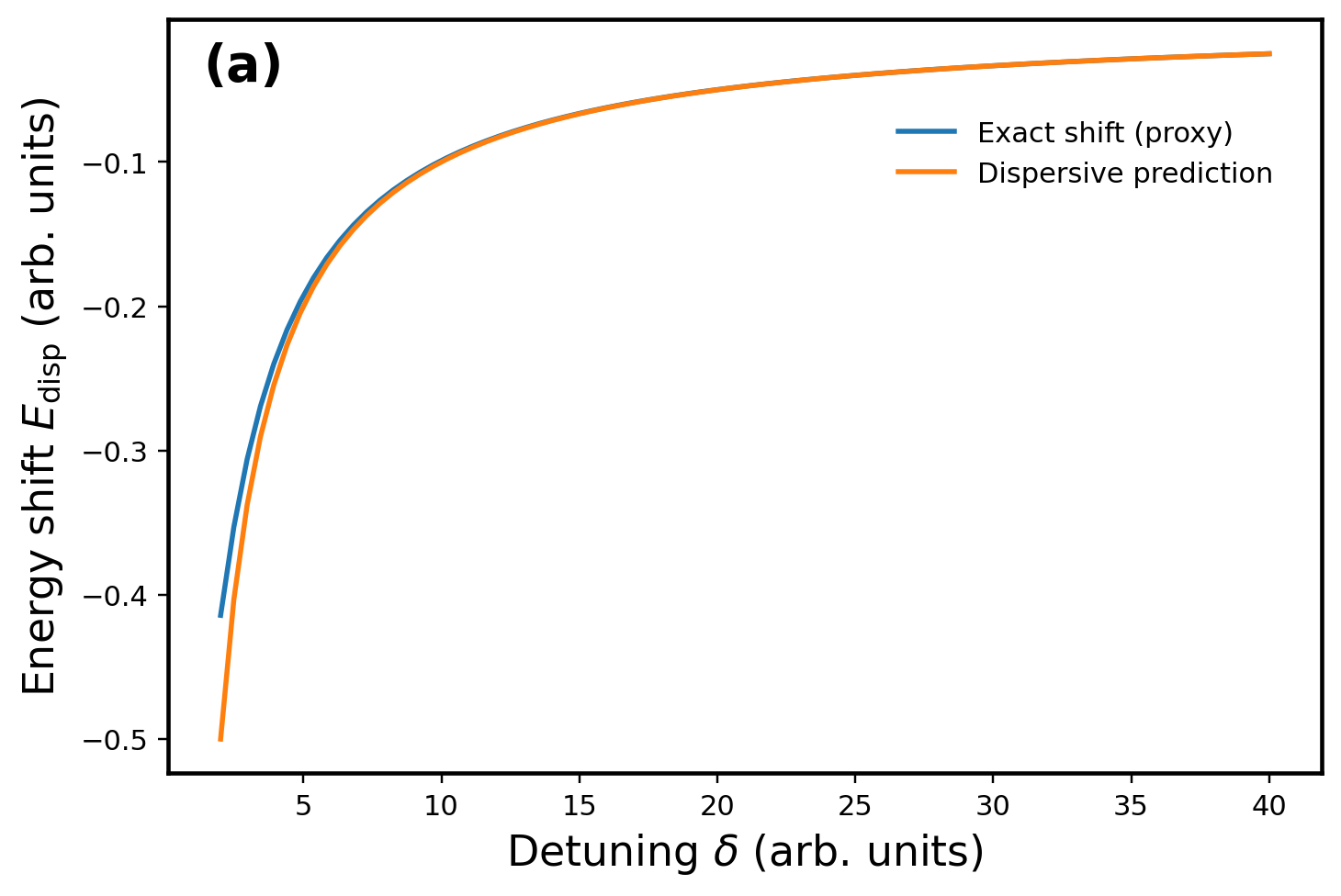}\hfill
\includegraphics[width=0.29\textwidth]{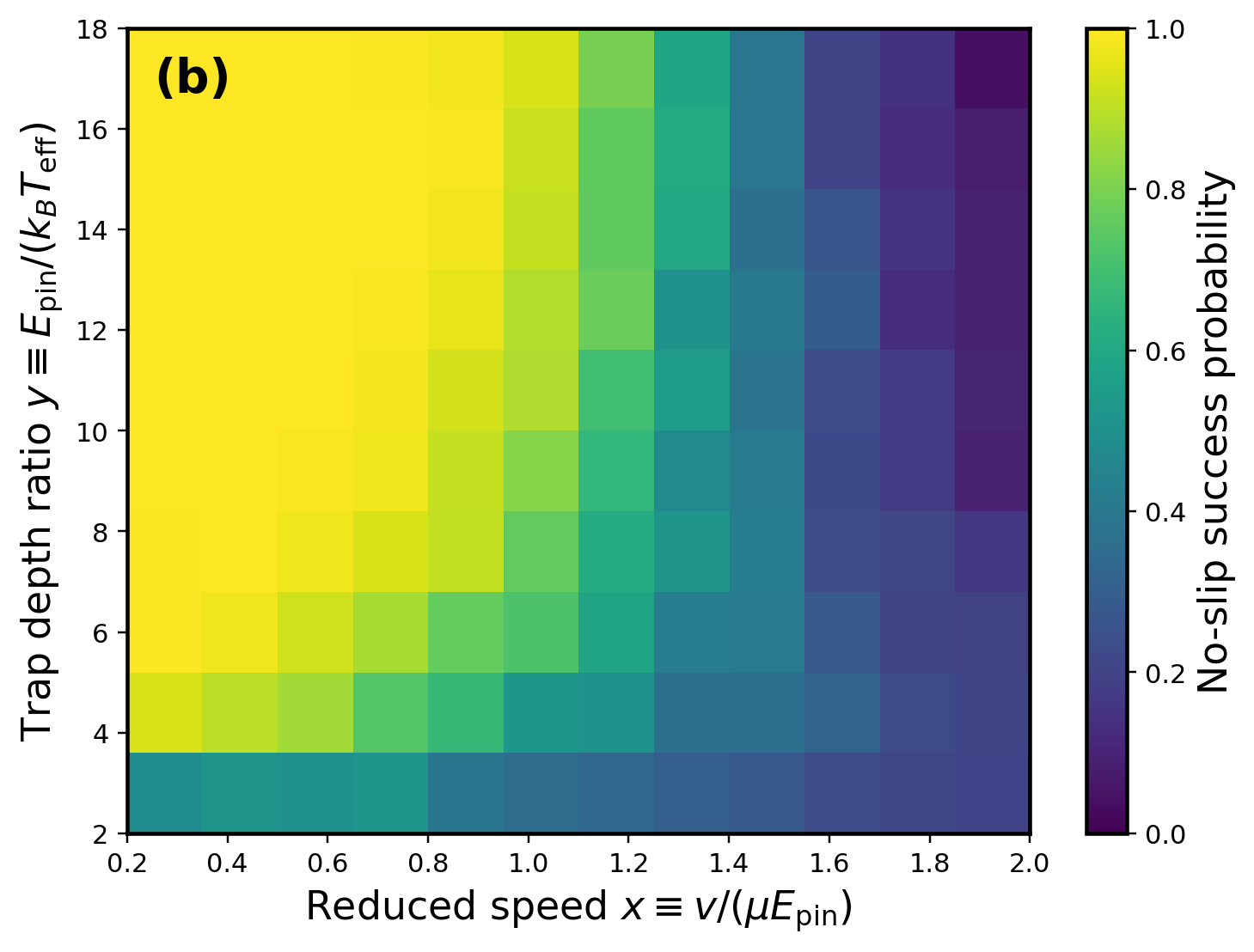}\hfill
\includegraphics[width=0.32\textwidth]{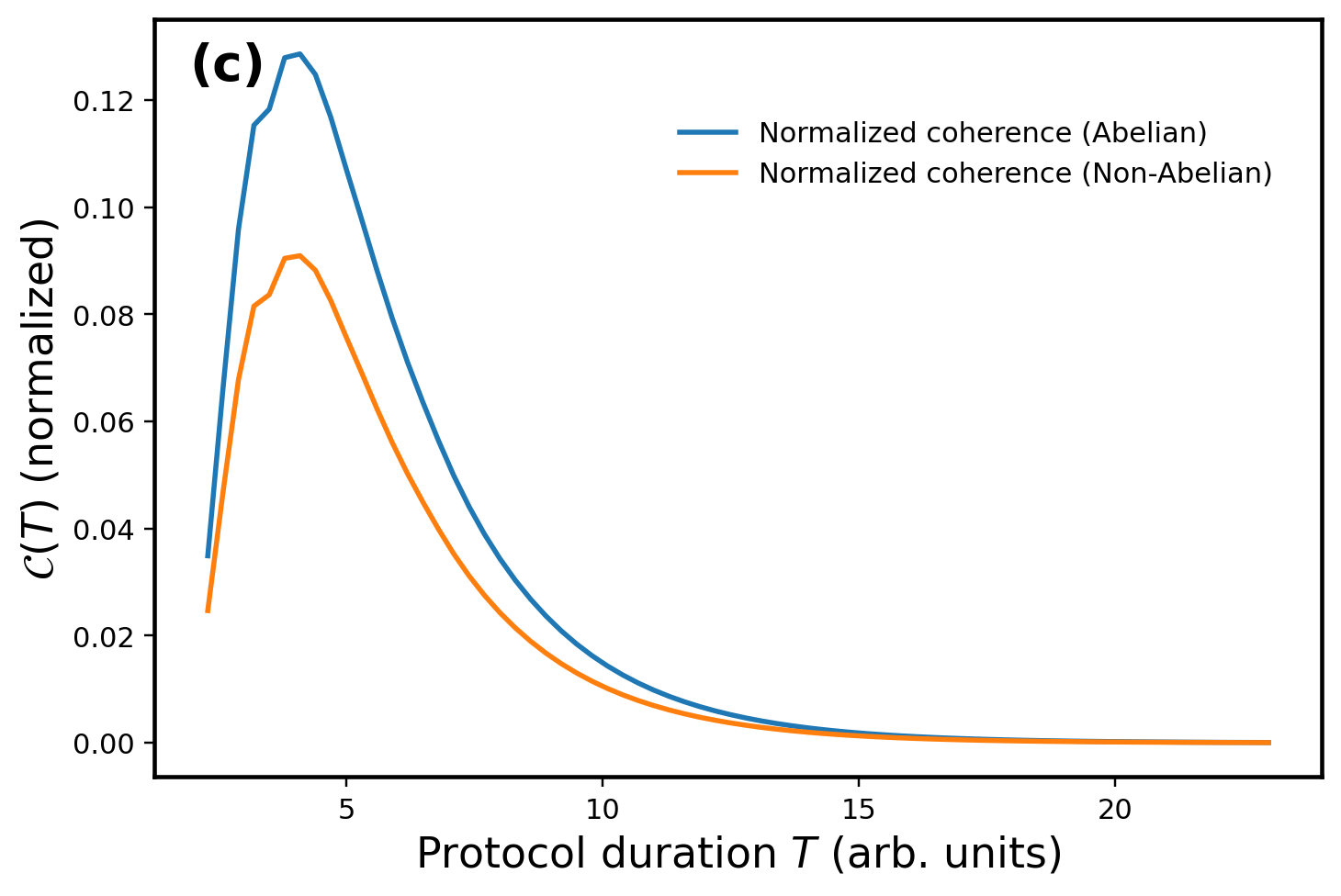}
\caption{\textbf{Appendix phenomenological proxy diagnostics.}
\textbf{(a)} Minimal proxy benchmark of dispersive elimination in a two-level model
(see Appendix Sec.~\ref{sec:app_dispersive_proxy}).
The plotted ``energy shift'' denotes the static dispersive level shift $E_{\mathrm{disp}}$, not the rotating pinning
amplitude $E_{\mathrm{pin}}$ used in panels (b,c). \textbf{(b)} Transport-locking sweep in the rotating pinning
landscape. The axes are rotation speed $v=2\pi/T$ and trap depth $E_{\mathrm{pin}}$ (arb.\ units), corresponding
directly to the dimensionless parameters $x\equiv v/(\mu E_{\mathrm{pin}})$ (horizontal, reversed) and
$y\equiv E_{\mathrm{pin}}/(k_BT_{\mathrm{eff}})$ (vertical) used in
Table~\ref{tab:validation_metrics_SM}. The yellow region indicates the high-fidelity locking regime
($P_{\mathrm{succ}}\to 1$), defined by the no-phase-slip criterion Eq.~\eqref{eq:SM_success}. This panel is a
coarse-grained locking diagnostic, not a microscopic simulation of guiding-center braiding dynamics.
\textbf{(c)} Normalized coherence proxy $\mathcal{C}(T)$ [Eq.~\eqref{eq:SM_C_def}], showing a generic optimal cycle
duration set by the competition between improved locking at longer $T$ and cavity coherence loss at shorter $T$. The
two curves illustrate the effect of the topological trace factor
$|\mathrm{Tr}[\rho U_{\mathrm{top}}(\Gamma)^{-2}]|$, using hypothetical values for pedagogical contrast. In the
explicit four-quasihole Ising example of Sec.~\ref{sec:Physical_Setup_Conditional_Braiding_Concept},
$U_{\mathrm{top}}(\Gamma)^{-2}\propto\mathbb{1}$ and the curves would coincide; state-dependent contrast requires a
different path or a larger fusion space, as discussed there.}
\phantomsection\label{fig:validation_SM}
\end{figure*}

Figure~\ref{fig:validation_SM} summarizes three phenomenological proxy diagnostics:
(a) a dispersive-elimination benchmark [Appendix Sec.~\ref{sec:app_dispersive_proxy}],
(b) a transport-locking sweep with axes $(v,E_{\mathrm{pin}})$ corresponding to the dimensionless parameters
$x\equiv v/(\mu E_{\mathrm{pin}})$ and $y\equiv E_{\mathrm{pin}}/(k_BT_{\mathrm{eff}})$
[Appendix Sec.~\ref{sec:app_transport_model}],
and (c) the normalized coherence proxy $\mathcal{C}(T)$ [Eq.~\eqref{eq:SM_C_def}] combining transport and cavity decay.
Table~\ref{tab:validation_metrics_SM} lists the observables and representative acceptance criteria.
Numerical stability checks confirm that the locking boundaries are not timestep artifacts
($\Delta P_{\mathrm{succ}} \sim 10^{-3}$). To avoid extrapolation artifacts, the $T$ range shown in panel (c) is
chosen to remain within the $x$-grid domain used in panel (b), via $x=2\pi/(T\mu E_{\mathrm{pin}})$.

\begin{table}[!t]
\caption{Compact diagnostic checklist used in Fig.~\ref{fig:validation_SM}. Each row links a load-bearing assumption to the corresponding numerical diagnostic and a representative acceptance target (dimensionless).}
\label{tab:validation_metrics_SM}
\centering
\small
\setlength{\tabcolsep}{4pt}
\renewcommand{\arraystretch}{1.15}
\begin{tabularx}{\linewidth}{@{}p{0.38\linewidth}X@{}}
\toprule
Assumption (where used) & Diagnostic and acceptance target \\
\midrule
Dispersive elimination valid [Appendix Sec.~\ref{sec:app_dispersive_proxy}] &
Require $|g/\delta|\ll 1$ so the relative Stark-shift error $\epsilon_{\mathrm{disp}}\ll 1$ and leakage
$p_e\simeq (g/\delta)^2\ll 1$ (representative: $|\delta/g|\gtrsim 5$ as a conservative minimum;
Fig.~\ref{fig:validation_SM}(a) shows visually clearer convergence of the exact and approximate shifts closer to
$|\delta/g|\sim 15$). \\

No phase slips during transport [Appendix Sec.~\ref{sec:app_phase_slip}; Fig.~\ref{fig:validation_SM}(b)] &
No-slip success probability $P_{\mathrm{succ}}$ [Eq.~\eqref{eq:SM_success}] is high
(representative: $P_{\mathrm{succ}}\ge 0.9$, typically for
$y\equiv E_{\mathrm{pin}}/(k_BT_{\mathrm{eff}})\gtrsim 6$ and
$x\equiv v/(\mu E_{\mathrm{pin}})\lesssim 1$). \\

Interferometric visibility survives loss [Appendix Sec.~\ref{sec:app_visibility}; Fig.~\ref{fig:validation_SM}(c)] &
Choose $T$ near the maximum of the normalized proxy $\mathcal{C}(T)$ [Eq.~\eqref{eq:SM_C_def}], requiring
$\mathcal{C}(T)$ not to be exponentially small. \\
\bottomrule
\end{tabularx}
\end{table}

\section{Proxy model for dispersive elimination used in panel (a)}
\label{sec:app_dispersive_proxy}
To benchmark the emergence of a dispersive, nonresonant energy shift without substantial level mixing, we use the 
minimal two-level proxy Hamiltonian
\begin{equation}
H_{\mathrm{proxy}}=
\begin{pmatrix}
0 & g \\
g & \delta
\end{pmatrix},
\label{eq:SM_Hproxy}
\end{equation}
where $g$ is a coupling scale and $\delta$ is the detuning. Its exact eigenvalues are
\begin{equation}
\lambda_{\pm}
=
\frac{\delta \pm \sqrt{\delta^2+4g^2}}{2}.
\label{eq:SM_Hproxy_eigs}
\end{equation}
For the positive-detuning dispersive branch, with $|g/\delta|\ll 1$ and $\delta>0$, the lower dressed-state shift is
\begin{equation}
E_{\mathrm{disp}}\equiv \lambda_- \simeq -\frac{g^2}{\delta},
\end{equation}
while the admixture (leakage) scales as
\begin{equation}
p_e\simeq \left(\frac{g}{\delta}\right)^2.
\end{equation}
Panel (a) plots the exact shift $E_{\mathrm{disp}}=\lambda_-$ together with the corresponding eigenvectors of 
Eq.~\eqref{eq:SM_Hproxy}, thereby confirming these dispersive scalings and distinguishing $E_{\mathrm{disp}}$ from the 
rotating trap depth $E_{\mathrm{pin}}$.

\end{document}